\DeclareUnicodeCharacter{FB01}{fi}
\documentclass[smallextended]{svjour3}       % onecolumn (second format)
\smartqed  % flush right qed marks, e.g. at end of proof
\usepackage{graphicx}

\usepackage{latexsym}
\usepackage[square,sort,comma,numbers]{natbib}
\usepackage{graphicx}

\usepackage{amsthm,amsmath,amssymb}
\usepackage{mathrsfs}
\usepackage{bm}
\usepackage{booktabs}
\usepackage{color}
\usepackage{lineno}
\begin{document}

\title{Edge-Enhanced Dual Discriminator Generative Adversarial Network for Fast MRI with Parallel Imaging Using Multi-view Information%\thanks{Grants or other notes
}

\titlerunning{PIDD-GAN for Fast MRI}        % if too long for running head

\author{Jiahao Huang, Weiping Ding, Jun Lv, ~\\ Jingwen Yang, Hao Dong, Javier Del Ser, Jun Xia, Tiaojuan Ren, Stephen T. Wong, Guang Yang}

\authorrunning{Huang et al.} % if too long for running head

\institute{
            Jiahao Huang \at
            College of Information Science and Technology, Zhejiang Shuren University, Hangzhou, 310015, China \\
            National Heart and Lung Institute, Imperial College London, London, United Kingdom %\\ \email{j.huang21@imperial.ac.uk}  
            \and
            Weiping Ding \at
            School of Information Science and Technology, Nantong University, Nantong 226019, China %\\ \email{dwp9988@163.com}
            \and
            Jun Lv \at
            School of Computer and Control Engineering, Yantai University, Yantai 264005, China %\\ \email{ljdream0710@pku.edu.cn}
            \and
            Jingwen Yang \at
            Department of Prosthodontics, Peking University School and Hospital of Stomatology, Beijing, China %\\\email{jingwen.yang@foxmail.com}
            \and
            Hao Dong \at
            Center on Frontiers of Computing Studies, Peking University, Beijing, China %\\ \email{hao.dong@pku.edu.cn}
            \and
            Javier Del Ser \at
            TECNALIA, Basque Research and Technology Alliance (BRTA), 48160 Derio, Spain\\
            University of the Basque Country (UPV/EHU), 48013 Bilbao, Spain %\\ \email{javier.delser@tecnalia.com}
            \and
            Jun Xia \at
            Department of Radiology, Shenzhen Second People’s Hospital, The First Afﬁliated Hospital of Shenzhen University Health Science Center, Shenzhen, China % \\ \email{xiajun@email.szu.edu.cn}
            \and
            Tiaojuan Ren \at
            College of Information Science and Technology, Zhejiang Shuren University, Hangzhou,310015, China % \\ \email{rentj@zjsru.edu.cn}
            \and
            Stephen T. Wong \at
            Systems Medicine and Bioengineering Department, Houston Methodist Cancer Center and Departments of Radiology and Pathology, Houston Methodist Hospital, Weill Cornell Medicine, Houston, TX, 77030, United States %\\ \email{stwong@houstonmethodist.org}
            \and
            Guang Yang \at
            National Heart and Lung Institute, Imperial College London, London, United Kingdom \\
            Cardiovascular Research Centre, Royal Brompton Hospital, London, United Kingdom %\\ \email{g.yang@imperial.ac.uk}
            \and
            Send correspondence to T.J. Ren \email{rentj@zjsru.edu.cn} and G. Yang \email{g.yang@imperial.ac.uk}
            }

\date{Received: date / Accepted: date}

\maketitle

\begin{abstract}
In clinical medicine, magnetic resonance imaging (MRI) is one of the most important tools for diagnosis, triage, prognosis, and treatment planning. However, MRI suffers from an inherent slow data acquisition process because data is collected sequentially in \textit{k}-space.
In recent years, most MRI reconstruction methods proposed in the literature focus on holistic image reconstruction rather than enhancing the edge information.
This work steps aside this general trend by elaborating on the enhancement of edge information. Specifically, we introduce a novel parallel imaging coupled dual discriminator generative adversarial network (PIDD-GAN) for fast multi-channel MRI reconstruction by incorporating multi-view information. The dual discriminator design aims to improve the edge information in MRI reconstruction. One discriminator is used for holistic image reconstruction, whereas the other one is responsible for enhancing edge information. An improved U-Net with local and global residual learning is proposed for the generator. Frequency channel attention blocks (FCA Blocks) are embedded in the generator for incorporating attention mechanisms. Content loss is introduced to train the generator for better reconstruction quality. We performed comprehensive experiments on Calgary-Campinas public brain MR dataset and compared our method with state-of-the-art MRI reconstruction methods. Ablation studies of residual learning were conducted on the MICCAI13 dataset to validate the proposed modules. Results show that our PIDD-GAN provides high-quality reconstructed MR images, with well-preserved edge information. The time of single-image reconstruction is below 5ms, which meets the demand of faster processing.

\end{abstract}

\section{Introduction}

% MRI
Magnetic resonance imaging (MRI) is one of the most important clinical tools for diagnosis, triage, prognosis, and treatment planning. MRI produces accurate, potentially high-resolution, and reproducible images with various contrast and functional information. Furthermore, it is non-invasive and harmless to the human body. However, MRI has an inherently slow data acquisition process since data is collected sequentially in \textit{k}-space, where speed is limited by physiological and hardware constraints, rather than directly in image space~\citep{Lustig}. Prolonged acquisition time can lead to severe motion artefacts due to patient movement and physiological motion. Early approaches to accelerate acquisition followed Nyquist-Shannon sampling criteria, e.g. implementing multiple radio frequency~\citep{Hennig} or gradient refocusing~\citep{Stehling}, with limited speed improvement. 

% CS & PI
Undersampling in \textit{k}-space can improve acquisition speed, but causes aliasing artefacts and blur. Several studies have attempted to reduce aliasing artefacts, such as parallel imaging (PI) and compressed sensing (CS). Parallel imaging was first introduced in 1997 as the simultaneous acquisition of spatial harmonic (SMASH)~\citep{Sodickson} to reduce scan time using multi-channel \textit{k}-space data and includes sensitivity encoding (SENSE)~\citep{Pruessmann} and generalized auto-calibrating partially parallel acquisition (GRAPPA)~\citep{Griswold}. The acceleration factor and geometry factor influence signal noise ratio for the reconstructed image, where geometry factor depends on the receiver coil distribution. Parallel imaging requires phased array coils where each coil receives data at the same time (i.e., parallel). If the local sensitivity for each coil is already known, then the field of view can be made arbitrarily small in the phase-encoding direction, enabling aliasing to be unwrapped using this information. 

Compressed sensing~\citep{D} reconstructs signals from significantly fewer measurements with higher sampling efficiency than traditional Nyquist-Shannon sampling. 
There are three requirements for successful CS~\citep{Ye}, which MRI naturally meets except for the third one:
(1) \emph{Transform sparsity}: MRI is compressible by transform coding and has a sparse representation in an appropriate transform domain.
(2) \emph{Incoherence}: It can be achieved by random undersampling MRI data in \textit{k}-space.
(3) \emph{Nonlinear reconstruction}: MRI should be reconstructed using a nonlinear method. The main task of CS is to find an appropriate nonlinear method to reconstruct MRI.
Several methods applying fixed sparsifying transforms for reconstruction have been proposed, e.g., total variation (TV)~\citep{Block}, curvelets~\citep{Beladgham}, and double-density complex wavelet~\citep{Zhu}, while adaptive sparse model represented by dictionary learning~\citep{S} was also developed as an extension.
In addition, CS-MRI worked well with parallel imaging data jointly.
Aelterman et al.~\citep{Aelterman} proposed a joint algorithm, i.e., COMPASS, of parallel imaging and compressed sensing for MRI reconstruction. Trzasko et al.~\citep{Trzasko} designed an offline, sparsity-driven reconstruction framework for Cartesian sampling time-series acquisitions.
However, the iterative computation significantly limits the speed of the reconstruction, leading to the fact that the traditional CS-MRI is hard to be applied in the clinical environment. 

% CNN 
Deep learning has been developed enormously recently. Convolutional neural networks (CNNs) have made a significant impact on many computer vision learning tasks, such as classification~\citep{Szegedy}, segmentation~\citep{Shelhamer}, object detection~\citep{Girshick}, and image reconstruction~\citep{C}. CNNs offer better feature extraction performance compared with traditional machine learning algorithms due to their deeper structure of hierarchically stacked neural layers. Consequently, several CNNs have been proposed for medical imaging to solve traditional limitations. CNN was first introduced for MRI reconstruction in~\citep{S2}, where it was used to identify mapping relationships between MR images obtained from zero-filled and fully-sampled \textit{k}-space data. 
Yang et al.~\citep{Yang} proposed a deep architecture, i.e., ADMM-Net, inspired by the alternating direction method of multipliers (ADMM) algorithm to optimizing CS based MRI models. 
Schlemper et al.~\citep{J} developed a deep cascade CNN to reconstruct dynamic sequences for 2D cardiac MR images from undersampled data.
Zhu et al.~\citep{Zhu2018} proposed a novel framework for MRI reconstruction, i.e., automated transform by manifold approximation (AUTOMAP), allowing a mapping between the sensor and the image domain to emerge from an appropriate corpus of training data.
Transfer learning has been also proposed to solve the data scarcity problem when training deep networks for accelerated MRI~\citep{Dar}.

% GAN
In 2014, Goodfellow et al.~\citep{Goodfellow} designed generative adversarial networks (GANs).
Then, various improved GAN models were presented.
Wasserstein GAN (WGAN)~\citep{Arjovsky} was proposed for improving the training stability of GAN and optimise the learning curves.
Radford et al.~\citep{Radford} introduced DCGAN, first applied CNNs to GANs, bridging the gap between supervised learning and unsupervised learning.
GAN-based models has been widely used in image-to-image translation~\citep{Zhu2, Choi,Xia2020,Pizzati2021}, super-resolution~\citep{Ledig2017, Chen2019, Jiang2021}, as well as MRI reconstruction~\citep{Yang2, Shaul2020, T}.

% GAN in MRI
Several studies have reported that the deep learning based GAN models perform well in MRI reconstruction.
% DAGAN
DAGAN~\citep{Yang2, Lv3} applied the modified U-Net~\citep{Ronneberger} architecture as the generator and incorporated additional perceptual loss by pre-trained VGG~\citep{Simonyan} networks. 
% KIGAN
Shaul et al.~\citep{Shaul2020, Lv3} introduced a dual-generator GAN, i.e., KIGAN, for both estimating the missing \textit{k}-space samples and refinement of the image space data.
% Recon/RefineGAN
Quan et al.~\citep{T, Lv3} presented RefineGAN, which was a variant of a fully-residual convolutional autoencoder and GAN with cyclic data consistency loss.
% PIC-GAN
Lv et al.~\citep{Lv}, proposed a deep GAN model, i.e., PIC-GAN, with parallel imaging for accelerated multi-channel MRI reconstruction, where data fidelity and regularisation terms were integrated into the generator. 
% GRAPPA-GANs
Nader et al.~\citep{Nader2021} combined the traditional MR reconstruction algorithm GRAPPA with a conditional GAN to build the GRAPPA-GAN, which was developed and tested on the multi-coil brain data.
% DAWGAN
Guo et al.~\citep{Guo2020} proposed DAWAGAN, which coupled WGAN with Recurrent Neural Networks to adopt the relationship among MRI slices for fast MRI reconstruction.
% Multi-Texture GAN
Hu et al.~\citep{Hu2021} designed a general texture-based GAN for MR image synthesis, where a multi-scale mechanism for the texture transfer between source and target domain was adopted.
% CSI-GAN
Ma et al.~\citep{Ma1} introduced a novel GAN-based model, i.e., CSI-GAN, for medical image enhancement, where illumination regularisation and structure loss were used as constraints of training.
% RE-Net
Zhang et al.~\citep{Zhang2} combined a Retinex model with the reverse edge attention network for cerebrovascular segmentation. The utilisation of reverse edge attention module significantly improved the performance of segmentation.
% SARA-GAN
Yuan et al.~\citep{Yuan1} incorporated the self-attention mechanism into the generator for a global understanding of images and improved the discriminator for the utilisation of prior knowledge.
% RSCA-GAN & HR
Li et al.~\citep{Li1} proposed RSCA-GAN for fast MRI reconstruction, where both spatial and channel-wise attention mechanisms were applied in the generator.
This team also applied the GAN-based model with attention mechanisms in the super-resolution task~\cite{Li2}.
% WPD-DAGAN
Chen et al.~\citep{Chen1}, incorporated wavelet packet decomposition into the de-aliasing GAN~\citep{Yang2} for the texture feature.
% PIGAN-Transfer learning
Lv et al.~\cite{Lv2} applied transfer learning to a parallel image coupled GAN model, improving the generalisability of networks based on small samples.
% FA-GAN
Jiang et al.~\cite{Jiang} proposed FA-GAN for the super-resolution task, where both global and local feature fusion were utilised in the generator for better performance. 
% ESSGAN & SOGAN
Zhou et al.~\cite{Zhou1} designed a GAN-based model, i.e., ESSGAN, with a structurally strengthened generator, which consisted of the strengthened connection and the residual in the residual block. This team also introduced a novel spatial orthogonal attention GAN model~\cite{Zhou2}, where the computational complexity was significantly decreased.

% GAN Focusing on Edge
Most proposed MRI reconstruction approaches focused on integral MR image property. The design of GAN-approaches relied on loss function definitions that do not consider structural characteristics of practical value that are present in the image such as edge information.
However, edge information can be crucially conclusive for clinical diagnosis. Accordingly, to solve this problem, studies related to edge information preservation in MRI reconstruction has been reported.
% Ea-GAN
Yu et al.~\citep{Yu2019} proposed Ea-GANs, in which the edge information was utilised via the Sobel operator. Ea-GANs contain a generator-induced gEa-GAN, and a discriminator-induced dEa-GAN, for enriching the reconstruction images with more details.
% EG GAN
Chai et al.~\citep{Chai2020} designed an edge-guided GAN (EG-GAN), to restore brain MRI images which decoupled reconstruction into edge connection and contrast completion.
% EDD GAN
Li et al.~\citep{Li2021}, proposed a dual-discriminator GAN, i.e., EDDGAN, of which one discriminator was used for holistic image reconstruction and the other one was for edge information preservation.

% multi-view
Recently, a large number of multi-view data based methods by considering the diversity of various views have been proposed~\citep{Wang1, Zhang1}. 
The main task of multi-view learning is to find a function to model each view and jointly optimises all the functions to improve the generalisation performance~\cite{Chao1, Zhao1}. 
In this work, the idea of multi-view learning was adopted. The multi-channel MR data we used in training was the multi-view information of the MRI raw data, by multi-coil parallel imaging. 
In addition, information from different views including image-space information, \textit{k}-space information and edge details were all included and used as constraints on the training process.

% Our Contribution
The methods mentioned above focused on the preservation of edge information from various aspects. However, all were based on single-channel MR data. In fact, complementary multi-view information can be provided by multi-channel MR data. This work takes advantage of parallel imaging technology by using multi-channel data rather than single-channel data.
In particular, we propose a novel parallel imaging coupled dual discriminator generative adversarial network (PIDD-GAN) for fast multi-channel MRI reconstruction. The main contributions can be summarised as follows.

\begin{itemize}
\item[$\bullet$] 
We introduce a dual discriminator GAN architecture, where two discriminators are used for holistic image reconstruction and edge information enhancement respectively.
\item[$\bullet$]  
The GAN generator is an improved U-Net with local and global residual learning that stabilises and accelerates the training process. Frequency channel attention blocks (FCA Blocks) are embedded in the improved U-Net to incorporate attention mechanisms.
\item[$\bullet$] 
Although the proposed network is designed for multi-channel MR image reconstruction, single-channel image reconstruction can also be accomplished.
\item[$\bullet$] 
Comprehensive experiments on the Calgary-Campinas public brain MR dataset \footnote{https://sites.google.com/view/calgary-campinas-dataset/} (359 subjects) compared the proposed method with current state-of-the-art MR reconstruction methods. Ablation studies for residual learning were conducted on the MICCAI13 dataset \footnote{https://mrbrains13.isi.uu.nl/} to validate the proposed modules. Experimental results confirm that the proposed PIDD-GAN achieves high reconstruction quality with faster processing time.
\end{itemize}

\section{Method}

\subsection{Traditional MRI}

The problem of traditional CS-MRI is to recover a vector $ x \in \mathbb{C}^N $ in image space from an undersampled vector $ y \in \mathbb{C}^M $ $(M \ll N)$ in \textit{k}-space, which can be expressed as follows

\begin{align}\label{formula:1}
y = \mathcal{M}\mathcal{F}x + n,
\end{align}

\noindent where $\mathcal{M}$ denotes the undersampling trajectory, $\mathcal{F}$ denotes the Fourier transform, and $n$ denotes noise. 

For parallel imaging, coil sensitivity encoding is incorporated in the reconstruction, i.e.,

\begin{align}\label{formula:2}
y=\mathcal{M}\mathcal{F}\mathcal{C}x + n, 
\end{align}

\noindent where $\mathcal{C}$ is the coil sensitivity. 

Let $\mathcal{E}$ be an operator including undersampling trajectory $\mathcal{M}$, Fourier transform $\mathcal{F}$ and coil sensitivity $\mathcal{C}$, the reconstruction problem can now be expressed as follows

\begin{align}\label{formula:3}
\mathop{\text{min}}\limits_{x} \frac{1}{2} 
\mid\mid y - \mathcal{E} x\mid\mid^2_2
+ \lambda R(x),
\end{align}

\noindent where $R(x)$ denotes regularisation terms on $x$, and $\lambda$ is a regularisation coefficient, controlling the degree of regularisation. $\mid\mid\cdot\mid\mid_2$ denotes the $l_2$ norm.

\subsection{Deep Learning-Based Fast MRI}

\subsubsection{CNN-Based Fast MRI}

A deep network can be incorporated into fast MRI reconstruction to generate image $f_{\mathrm{CNN}}(x_u\mid\theta)$ from the zero-filled image $x_u$ , where $\theta$ are the optimised parameters of the deep network. The problem can be expressed as follows

\begin{align}\label{formula:4}
\mathop{\text{min}}\limits_{x} \frac{1}{2} 
\mid\mid y - \mathcal{E} x\mid\mid^2_2
+ \lambda R(x)
+ \zeta \mid\mid x - f_{\mathrm{CNN}}(x_u \mid \theta) \mid\mid^2_2,
\end{align}

\noindent where $\lambda$ and $\zeta$ are coefficients that balance each term.

\subsubsection{GAN-Based Fast MRI}

In general, a GAN model consists of two parts: a generator and a discriminator. The generator aims to produce fake data $G_{\theta_G}(\bm{z})$ that is as real as possible by modelling and sampling the distribution of the ground truth $\bm{x}$, so that samples drawn from the modelled distribution succeed at deceiving the discriminator. The goal of the discriminator is to distinguish the fake data generated by the generator from the ground truth. 

Ideally, the best discriminator can be represented as 

\begin{align}\label{formula:5}
D_{\theta_D}(\bm{x})=1, \quad\quad D_{\theta_D}(G_{\theta_G}(\bm{z}))=0.
\end{align}

In this way, the generator and the discriminator form a min-max game. The training process of GAN can be described as follows

\begin{align}\label{formula:6}
\mathop{\text{min}}\limits_{\theta_G} 
\mathop{\text{max}}\limits_{\theta_D}
\mathcal{L}(\theta_G, \theta_D)
=\mathbb{E}_{\bm{x} \sim p_{\mathrm{data}}(\bm{x})}
[\mathop{\text{log}} D_{\theta_D}(\bm{x})]
+\mathbb{E}_{\bm{z} \sim p_{\bm{z}}(\bm{z})}
[\mathop{\text{log}} (1-D_{\theta_D}(G_{\theta_G}(\bm{z})))],
\end{align}

\noindent where $p_{\mathrm{data}}(\bm{x})$ is the distribution of real data, and $p_{\bm{z}}(\bm{z})$ is the latent variables distribution. 
During the training process, both sides constantly optimise themselves until the balance is reached --- neither side can get better, that is, the fake samples are completely indistinguishable from the true samples.

However, the discriminator may be very confident and almost always output 0 initially. To circumvent this problem practically, the loss function is replaced by:

\begin{align}\label{formula:7}
\mathop{\text{min}}\limits_{\theta_G} 
\mathop{\text{max}}\limits_{\theta_D}
\mathcal{L}(\theta_G, \theta_D)
=\mathbb{E}_{\bm{x} \sim p_{\mathrm{data}}(\bm{x})}
[\mathop{\text{log}} D_{\theta_D}(\bm{x})]
-\mathbb{E}_{\bm{z} \sim p_{\bm{z}}(\bm{z})}
[\mathop{\text{log}} (D_{\theta_D}(G_{\theta_G}(\bm{z})))].
\end{align}

When the GAN model is used for the reconstruction task, the generator is trained to generate reconstructed MR images $G_{\theta_G}(x_u)$ from zero-filled undersampled MR images $x_u$. The discriminator is trained to distinguish $G_{\theta_G}(x_u)$ from the ground truth MR image $x_t$ by maximizing the log-likelihood for estimating the conditional probability, which can be represented as

\begin{align}\label{formula:8}
D_{\theta_D}(x_t)=1, \quad\quad D_{\theta_D}(G_{\theta_G}(x_u))=0.
\end{align}

The adversarial loss $\mathcal{L}_{\mathrm{adv}}$ that drives the training process can be parameterised by $\theta_G$ and $\theta_D$ as follows

\begin{align}\label{formula:9}
\mathop{\text{min}}\limits_{\theta_G} 
\mathop{\text{max}}\limits_{\theta_D}
\mathcal{L}_{\mathrm{adv}}(\theta_G, \theta_D)
=\mathbb{E}_{{x_t} \sim p_{\mathrm{train}}({x_t})}
[\mathop{\text{log}} D_{\theta_D}({x_t})]
-\mathbb{E}_{{x_u} \sim p_G({x_u})}
[\mathop{\text{log}} (D_{\theta_D}(G_{\theta_G}({x_u}))].
\end{align}

\subsection{Proposed Dual Discriminator GAN for Fast MRI Reconstruction}

\subsubsection{Formulation}

In this work, a first discriminator $D_1$ is used to distinguish the reconstructed MR images $\hat x_u$ from the ground truth MR images $x_t$, whereas an additional discriminator $D_2$ is designed to assist the reconstruction of the edge information. 
Edge information is usually extracted by means of a Sobel operator $\mathcal{S}(\cdot)$. The edge information of the reconstructed MR image $\mathcal{S}({\hat x_u})$ and the edge information of the MR ground truth $\mathcal{S}({x_t})$ are fed to $D_2$, so that the result is counted into the adversarial loss $\mathcal{L}_{\mathrm{adv}}$. The new adversarial loss $\mathcal{L}_{\mathrm{adv}}$ can be defined as follows

\begin{align}\label{formula:10}
&\mathop{\text{min}}\limits_{\theta_G} 
\mathop{\text{max}}\limits_{\theta_{D_1}}
\mathop{\text{max}}\limits_{\theta_{D_2}}
\mathcal{L}_{\mathrm{adv}}(\theta_G, \theta_{D_1}, \theta_{D_2})\nonumber
\\&=\mu\{\mathbb{E}_{{x_t} \sim p_{\mathrm{train}}({x_t})}
[\mathop{\text{log}} D_{\theta_{D_1}}({x_t})]
-\mathbb{E}_{{x_u} \sim p_G({x_u})}
[\mathop{\text{log}} D_{\theta_{D_1}}(\hat x_u)]\}\nonumber
\\&+\nu\{\mathbb{E}_{{x_t} \sim p_{\mathrm{train}}({x_t})}
[\mathop{\text{log}} D_{\theta_{D_2}}(\mathcal{S}({x_t}))]
-\mathbb{E}_{{x_u} \sim p_G({x_u})}
[\mathop{\text{log}} D_{\theta_{D_2}}(\mathcal{S}(\hat x_u))]\} ,
\end{align}

\noindent where $\mu$ and $\nu$ denote the weights of the discriminator for the holistic image and the edge information correspondingly, and $\hat x_u = G_{\theta_G}({x_u})+x_u$ as it will be later discussed.

\subsection{Network Architecture}

\begin{figure}[htbp]
\centering
\includegraphics[width=12cm]{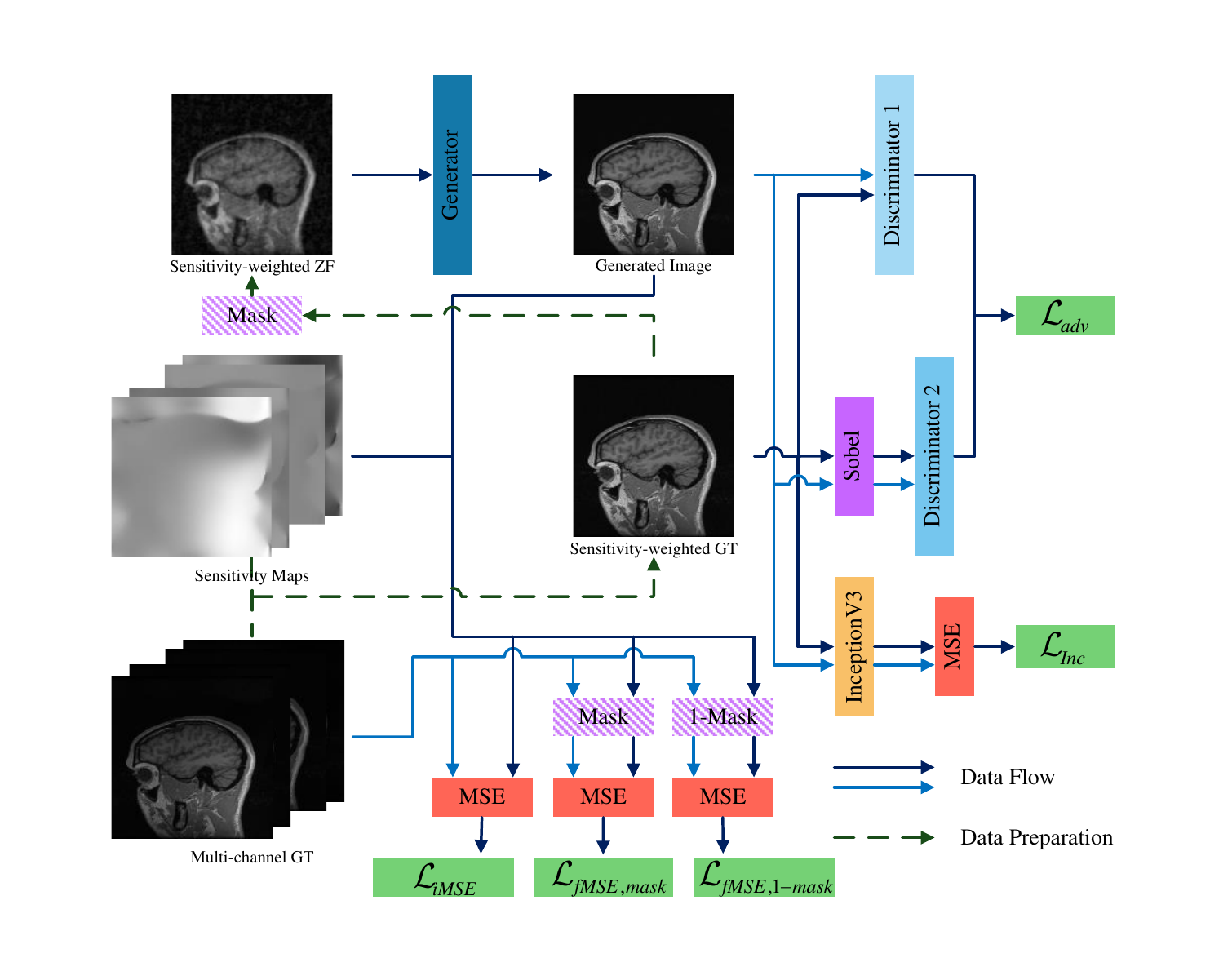}
\caption{The architecture of our PIDD-GAN. The generator produces generated MR images from sensitivity-weighted zero-filled MR images. Sensitivity maps and multi-channel ground truth images are used to produce sensitivity-weighted ground truth images. Two discriminators are used for the reconstruction of holistic images and edge information respectively. A pre-trained Inception V3 model is applied for the perceptual loss. Dark blue and sky blue lines are both used to represent the data flow for easier identification. Green dash lines are used to represent the data preparation steps before the training.
(MASK: undersampling trajectory mask, 1-MASK: inverse undersampling trajectory mask, InceptionV3: a pre-trained InceptionV3 model, MSE: mean squared error, Sobel: Sobel operator, Discriminator 1: the discriminator for holistic images, Discriminator 2: the discriminator for edge information).}
\label{fig:pidd_gan}
\end{figure}

The overall architecture of the proposed PIDD-GAN for MR image reconstruction is shown in Fig.~\ref{fig:pidd_gan}. Sensitivity-weighted ground truth $x_t$ is derived from the multi-channel ground truth $x_t^q$, corresponding sensitivity map $\mathcal{C}^q$ ($q$ denotes the coil number) and sensitivity-weighted zero-filled image $x_u$ by undersampling $x_t$. The generator produces the reconstructed MR image $\hat x_u$ from $x_u$. Two discriminators are used for the holistic image and the edge information, respectively. 

Ronneberger et al.~\citep{Ronneberger} proposed U-Net for semantic segmentation of medical images. This deep architecture comprises an encoder path to capture context and a symmetric decoder path that enables precise localisation, i.e., a U-shaped model. Skip connections are applied between corresponding layers in encoder and decoder paths, passing features directly from undersampling path to upsampling path. U-Net can be trained in an end-to-end manner and performs well in medical image segmentation~\citep{Yang2, T, Li2021}. 

\begin{figure}[htbp]
\centering
\includegraphics[width=12cm]{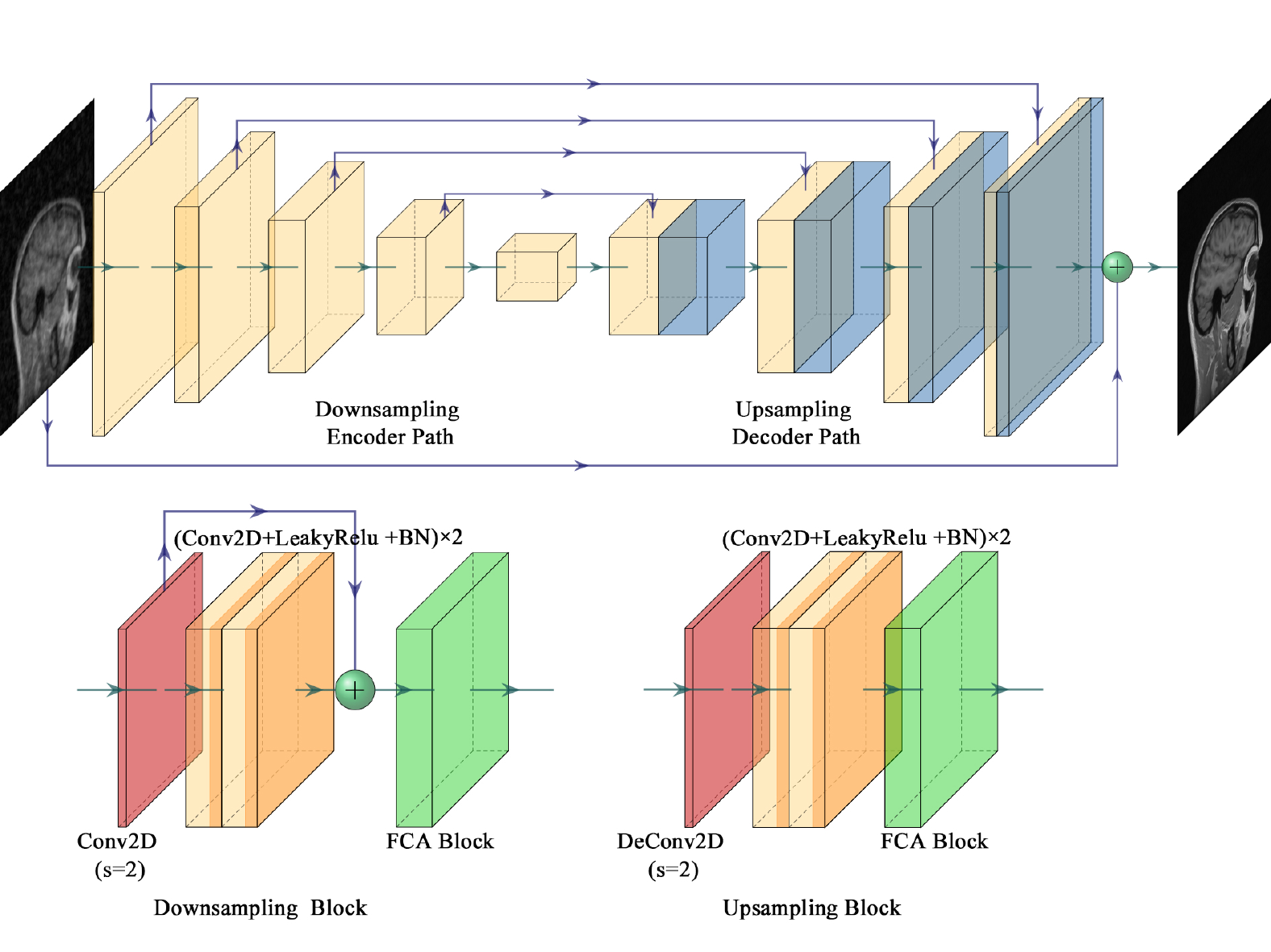}
\caption{The structure of the generator in PIDD-GAN. Four cascaded downsampling blocks and four cascaded upsampling blocks are placed in encoder and decoder paths respectively. Skip connection and concatenation are used between layers with the same scales. A shortcut connection is applied between the input and output of the generator for further refined learning. (Conv2D: 2D convolutional layer, DeConv2D: 2D deconvolution layer, BN: batch normalisation, LeakyRelu: Leaky ReLU layer.}
\label{fig:generator}
\end{figure}

An improved U-Net is proposed as the generator in our proposed PIDD-GAN for higher reconstruction performance, where the generator input is sensitivity-weighted zero-filled image $x_u$. As shown in Fig.~\ref{fig:generator}, in our improved U-Net, cascaded downsampling blocks are placed in the encoder path (left), and corresponding cascaded upsampling blocks are placed in the decoder path (right). Skip connection and concatenation are applied between the downsampling blocks and the symmetrical upsampling blocks with the same scale to preserve the feature from different levels and, ultimately, yield better reconstruction details. 

Fig.~\ref{fig:generator} shows the $i^{ \rm th}$ downsampling block structure. First, a $3 \times 3$ convolution layer with stride = 2 is applied to downsample the $(i-1)^{ \rm th}$ output. Then, a residual block extracts further features and avoids gradient vanishing and exploding problems~\citep{He}. There are two $3 \times 3$ convolution layers in the backbone and a $1 \times 1$ convolution layer adjusting the channel and fusing feature maps at different scales. Leaky ReLU layers and Batch normalisation (BN) are applied after each convolution layer except the final one. Finally, a FCA Block~\citep{Qin} learns the different channel weights with attention. 
The structure of upsampling blocks and downsampling blocks are similar. A deconvolution layer with stride = 2 is applied to upsample the $(i-1)^{ \rm th}$ output. The shortcut is removed to reduce computation cost but maintain high reconstruction quality here.

Traditional CNNs treat all channels in a feature map with the same importance, ignoring the importance differences. Therefore, the attention mechanism is adopted to make use of importance difference information by learning the different channel weights, i.e., effective channels have high weights and ineffective channels have small weights, which helps to train the model and enhance the results. 

\begin{figure}[htbp]
\centering
\includegraphics[width=12cm]{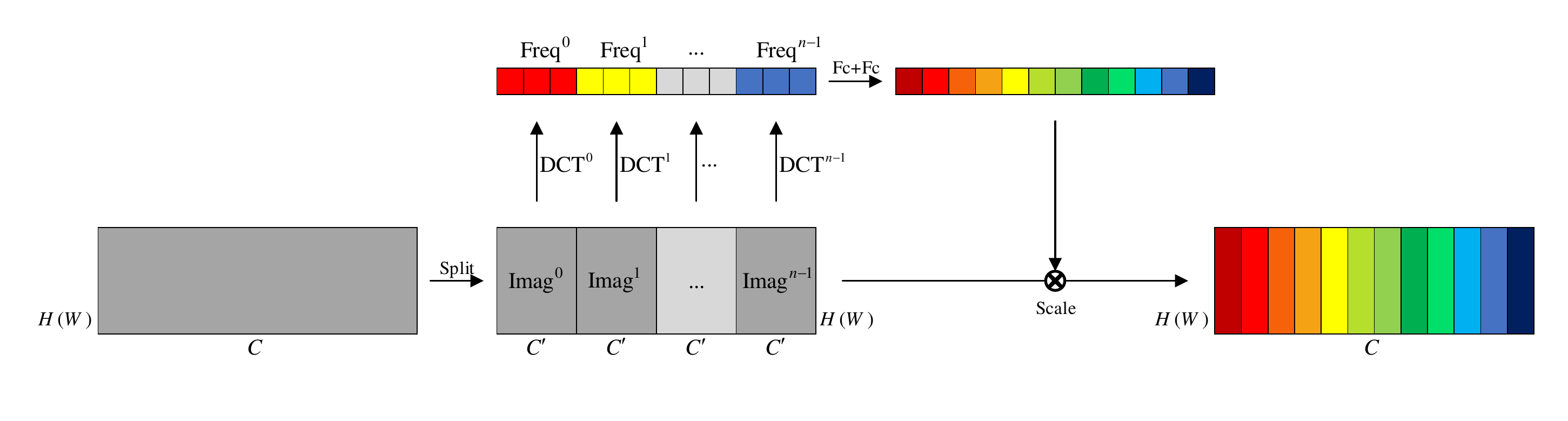}
\caption{The structure of the FCA Block. An $H \times W \times C$ feature map is divided into $n$ $H \times W \times C^{\prime}$ parts, i.e., $\text{Imag}^{i}\quad(i=0,1...n-1)$. The $i^{ \rm th}$ frequency component $\text{Freq}^{i}$ is obtained by corresponding $\text{Imag}^{i}$ using $\text{DCT}^{i}$. The weight of channels can be counted by all the frequency components using two full connection layers (FC in the Fig).}
\label{fig:fca_block}
\end{figure}

FCA Block~\citep{Qin} is a novel attention mechanism based on the squeeze and excitation block (SE Block)~\citep{Hu}, as shown in Fig.~\ref{fig:fca_block}. First, an $H \times W \times C$ feature map is squeezed into a $1\times C$ vector. Then channel weights are extracted by two fully-connected layers, and channel weights multiply the original feature maps. A two-dimensional discrete cosine transform (DCT) is applied in FCA Blocks to squeeze the feature map, rather than global average pooling employed in SE Blocks, since this latter operation is equivalent to the lowest DCT frequency. Hence using only GAP leads to loss of other frequency components in the feature channel containing useful information. In the squeezing step, the $H \times W \times C$ feature map is divided evenly into $n$ parts (each size is $H \times W \times C^{\prime}$). The squeezing step can be represented as 

\begin{align}\label{formula:11}
\text{Freq}^{i}=\text{DCT}^{i}(\text{Imag}^{i}), \quad\quad(i=0,1...n-1),
\end{align}

\noindent where $\text{DCT}^{i}$ denotes the $i^{ \rm th}$ preset DCT template and $\text{Freq}^{i}$ denotes the corresponding frequency component. In this work, FCA Blocks are utilised after every residual blocks in the generator.
 
As the network architecture goes deeper, more granular information can be extracted, but leading to gradient vanishing and exploding problems, making the network converge slowly. The introduction of residual learning~\citep{He} solves this problem effectively. The main idea of residual learning is the utilisation of shortcut connections between the convolution layer. It makes the deep network easier to be trained and converge faster. 
Global residual learning (GR) is applied in our improved U-Net. The output of the generator adopts $\hat x_u = G_{\theta_G}({x_u})+x_u$ instead of $\hat x_u = G_{\theta_G}({x_u})$ in the original U-Net. This change transfers the generator from a conditional generative function to a refinement function. 
This work also applies local residual learning (LR) by the shortcut connection in each residual block in the downsampling path. The utilisation of LR aimed to stabilise the training and accelerate the model convergence. 

Traditional GAN trains a single discriminator to compete against the generator. Although it improves reconstruction quality compared with other methods, only integral MR image properties are considered, without enhancing edge details. The current study proposes a dual discriminator GAN for edge information enhancement. The generated MR image $\hat x_u$ and sensitivity-weighted MR ground truth $x_t$ are fed into discriminator $D_1$ for holistic image reconstruction. We use the Sobel operator $\mathcal{S}(\cdot)$ to extract the edge information from MR images, and input edge information for the reconstructed image $\mathcal{S}(\hat x_u)$ and ground truth $\mathcal{S}(x_t)$ into $D_2$. Thus, both holistic image information and edge details can be simultaneously reconstructed.

\begin{figure}[htbp]
\centering
\includegraphics[width=12cm]{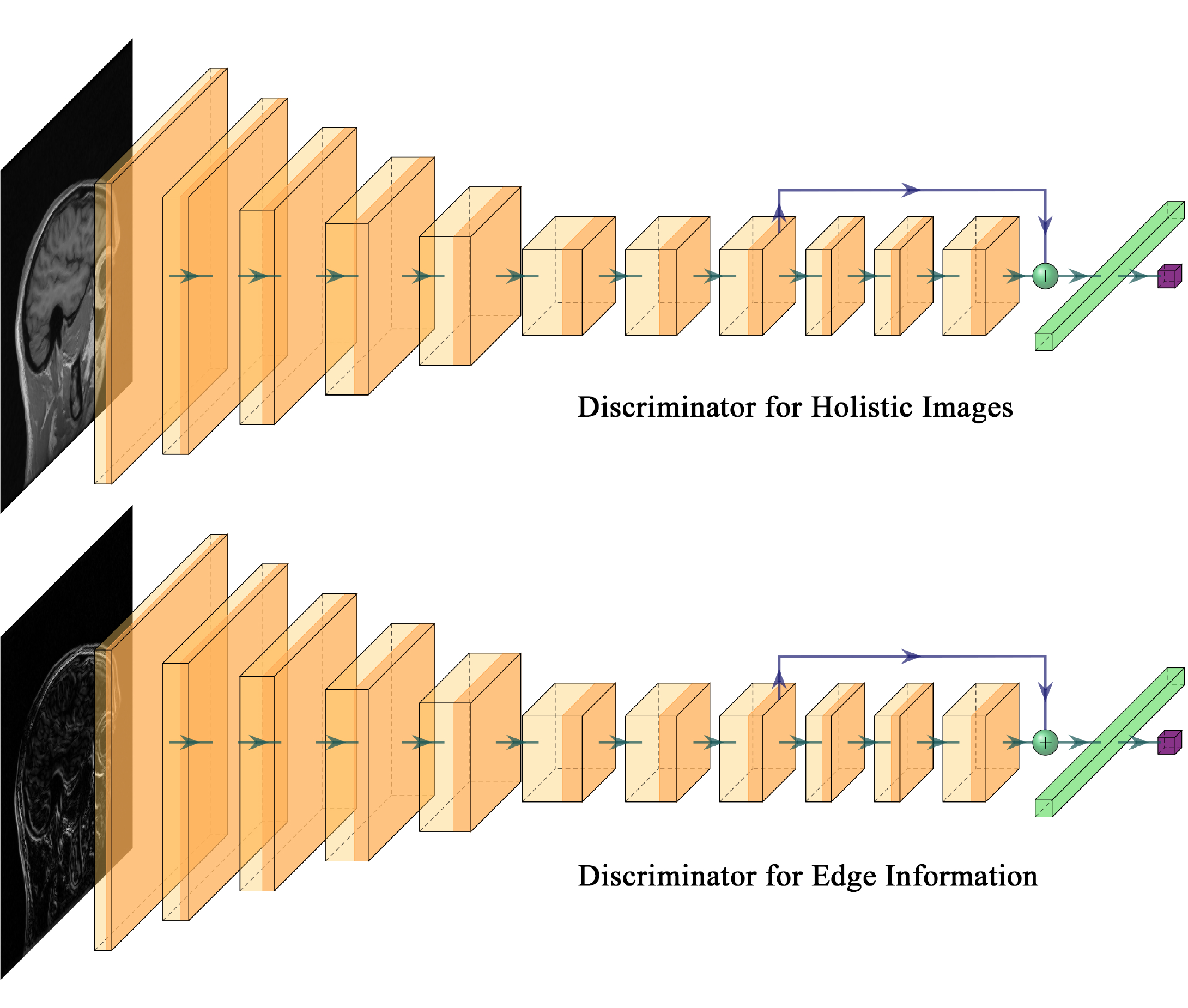}
\caption{The structure of the discriminator in PIDD-GAN. A standard 11-layer CNN is used as the discriminator, where each convolutional layer is followed by a BN layer and a Leaky ReLU layer. A full connection layer and a Sigmoid layer are cascaded at the end of discriminators to output the result of classification. The upper discriminator is the discriminator for holistic image and the lower one is for edge information.}
\label{fig:discriminator}
\end{figure}

Fig.~\ref{fig:discriminator} shows the common network structure used for both discriminators. First, a cascade of $3 \times 3$ convolution layers with stride = 2 downsamples and extracts MR image features. Two $1 \times 1$ convolution layers and a residual block follow the final $3 \times 3$ convolution layer. The residual block consists of three $1 \times 1$ convolution layers, and input and output are connected by a shortcut. All convolution layers above are followed by a BN layer and Leaky ReLU layer. Finally a full connection layer and a Sigmoid layer output the prediction results. Results of both discriminators are incorporated into the adversarial loss $\mathcal{L}_{\mathrm{adv}}$.

\subsection{Loss Function}

This study introduces content loss to train the generator for better reconstruction quality. Content loss comprises pixel-wise mean square error (MSE) loss $\mathcal{L}_{\mathrm{iMSE}}$; the frequency MSE loss $\mathcal{L}_{\mathrm{fMSE,mask}}$ and $\mathcal{L}_{\mathrm{fMSE,1-mask}}$; and the perceptual InceptionV3 loss $\mathcal{L}_{\mathrm{Inc}}$. Pixel-wise MSE loss can be defined as 

\begin{align}\label{formula:12}
\mathop{\text{min}}\limits_{\theta_G}  
\mathcal{L}_{\mathrm{iMSE}}(\theta_G) =
\sum_q
\frac{1}{2}\mid\mid x_t^q - \mathcal{C}^q \hat x_u \mid\mid^2_2,
\end{align}

\noindent where $\mathcal{C}^q$ is the sensitivity map of the $q^{ \rm th}$ coil. $\mathcal{L}_{\mathrm{iMSE}}$ is used to reduce the artefact between the generated image and the ground truth. However, optimisation with only $\mathcal{L}_{\mathrm{iMSE}}$ would make the reconstructed image lack coherent image details. Therefore, frequency MSE loss is applied to train the generator in \textit{k}-space.

The frequency MSE loss can be defined as

\begin{align}\label{formula:13}
\mathop{\text{min}}\limits_{\theta_G}  
\mathcal{L}_{\mathrm{fMSE,mask}}(\theta_G) =
\sum_q
\frac{1}{2}\mid\mid y_{\mathcal{M}}^q - \mathcal{M}\mathcal{F}\mathcal{C}^q \hat x_u \mid\mid^2_2,
\end{align}

\begin{align}\label{formula:14}
\mathop{\text{min}}\limits_{\theta_G}  
\mathcal{L}_{\mathrm{fMSE,1-mask}}(\theta_G) =
\sum_q
\frac{1}{2}\mid\mid y_{1-\mathcal{M}}^q - (1-\mathcal{M})\mathcal{F}\mathcal{C}^q \hat x_u \mid\mid^2_2,
\end{align}

\noindent where $\mathcal{L}_{\mathrm{fMSE,mask}}$ eliminates differences between undersampled generated images $\mathcal{M}\mathcal{F}\mathcal{C}^q \hat x_u$ and undersampled \textit{k}-space measurements $y_{\mathcal{M}}^q$. $\mathcal{L}_{\mathrm{fMSE,1-mask}}$ is used to minimise the differences between the interpolated data based on the generated image $(1-\mathcal{M})\mathcal{F}\mathcal{C}^q \hat x_u$ and the unacquired \textit{k}-space data $y_{1-\mathcal{M}}^q$.

In addition, the perceptual Inception V3 loss can be defined as 

\begin{align}\label{formula:15}
\mathop{\text{min}}\limits_{\theta_G}  
\mathcal{L}_{\mathrm{Inc}}(\theta_G) =
\frac{1}{2}\mid\mid f_{\mathrm{Inc}}(x_t) - f_{\mathrm{Inc}}(\hat x_u) \mid\mid^2_2 ,
\end{align}

\noindent where $f_{\mathrm{Inc}}(\cdot)$ denotes the Inception V3 network \citep{Szegedy}. $\mathcal{L}_{\mathrm{Inc}}$ is used to optimise the perceptual quality of reconstructed results.

The adversarial loss is defined as 

\begin{align}\label{formula:16}
&\mathop{\text{min}}\limits_{\theta_G} 
\mathop{\text{max}}\limits_{\theta_{D_1}}
\mathop{\text{max}}\limits_{\theta_{D_2}}
\mathcal{L}_{\mathrm{adv}}(\theta_G, \theta_{D_1}, \theta_{D_2})\nonumber
\\&=\mu\{\mathbb{E}_{{x_t} \sim p_{\mathrm{train}}({x_t})}
[\mathop{\text{log}} D_{\theta_{D_1}}({x_t})]
-\mathbb{E}_{{x_u} \sim p_G({x_u})}
[\mathop{\text{log}} D_{\theta_{D_1}}(\hat x_u)]\}\nonumber
\\&+\nu\{\mathbb{E}_{{x_t} \sim p_{\mathrm{train}}({x_t})}
[\mathop{\text{log}} D_{\theta_{D_2}}(\mathcal{S}({x_t}))]
-\mathbb{E}_{{x_u} \sim p_G({x_u})}
[\mathop{\text{log}} D_{\theta_{D_2}}(\mathcal{S}(\hat x_u))]\} ,
\end{align}

Hence, the total loss can be described as

\begin{align}\label{formula:17}
\mathcal{L}_{\mathrm{TOTAL}} 
= \alpha \mathcal{L}_{\mathrm{iMSE}} 
+ \beta (\mathcal{L}_{\mathrm{fMSE,mask}} +  \mathcal{L}_{\mathrm{fMSE,1-mask}}) 
+ \gamma \mathcal{L}_{\mathrm{Inc}} 
+ \mathcal{L}_{\mathrm{adv}} ,
\end{align}

\noindent where $\alpha$, $\beta$ and $\gamma$ are coefficients balancing each term in the loss function.

\section{Experiments and Results}

\subsection{Datasets}

This work used the Calgary Campinas multi-channel dataset~\citep{Souza2018} and the MICCAI 2013 grand challenge single-channel dataset~\citep{mendrik2015mrbrains}, which are both publicly available for the experiment section.

The Calgary Campinas dataset was used to train and validate our proposed method and compare with other methods. The MR images were acquired with a 12-channel coil. We randomly chose 15360 12-channel T1-weighted brain MR images. The dataset was divided into training, validation, and testing sets (7680, 3072, and 4608 slices respectively), according to the ratio of 5:2:3.

The MICCAI 2013 grand challenge dataset was used for ablation studies. We randomly chose 18850 single-channel T1-weighted brain MRI images, and divided these into training, validation, and testing sets (9935, 3974, and 5961 slices, i.e., ratio of 5:2:3, respectively).

\subsection{Implementation Detail}

The proposed PIDD-GAN was implemented using PyTorch, and trained and tested on an NVIDIA TITAN RTX GPU with 24GB GPU memory. We set the same hyperparameters for all experiments. Adam optimiser with a momentum of 0.5 was adopted during training. Empirically, we set $\alpha = 15$, $\beta = 0.1$, $\gamma=10$ (for Inception V3) or $0.0025$ (for VGG) in the total loss function, and $\mu = 0.6$, $\nu = 0.4$ for discriminator weights. Initial and minimal learning rates were 0.001 and 0.00001, decayed by $50\%$ every 5 epochs. The batch size was set to 12. An early stopping mechanism was adopted to halt training and prevent overfitting: training was stopped if there were no normalised mean square error (NMSE) reduction on the validation set for 8 epochs.

To evaluate the proposed method, we compared the following variations: (1) PIDD: dual discriminator GAN model trained with parallel imaging dataset; (2) PISD: single discriminator GAN trained with parallel imaging dataset; (3) nPIDD: dual discriminator GAN model trained with single-channel imaging dataset. Thus, the role for each component in the proposed method can be compared more fairly and clearly.

\subsection{Evaluation Methods}

Various assessment methods were applied to evaluate reconstruction quality. Normalised mean square error measures average squared difference between generated images and ground truth. Structural similarity (SSIM) was used to measure similarity between two images and hence predict perceived quality. Peak signal-to-noise ratio (PSNR) is the ratio between maximum signal power and corrupting noise power, which quantifies the representations fidelity.

However, PSNR and SSIM do not necessarily correspond with visual quality for human observers. Therefore, we adopted two different metrics to evaluate reconstruction quality. Fréchet inception distance (FID)~\citep{Heusel} measures similarity between two sets of images, calculated by computing the Fréchet distance between two Gaussian fitted feature representations for the inception network. FID correlates well with human derived visual quality and it is widely used to evaluate GAN sample quality.

Meanwhile, the judgement of domain experts is considered. Mean opinion score (MOS) from expert observers was used to evaluate the holistic image and edge information for the reconstructed images.
Likert scales~\citep{Yang3} from 1 (poor), 2 (fair), 3 (good) to 4 (very good) were used that were based on the holistic image quality and edge information quality, the visibility of the atrial scar and occurrence of the artefacts. 
All compared reconstruction results of different methods were randomly shuffled and blinded for the expert observers scrutinisation. 

\subsection{Comparisons with Other Methods}

To better evaluate the reconstruction performance of the PIDD-GAN, we compare it with other traditional and MR reconstruction algorithms, including TV~\citep{Block}, L1-EPSRiT~\citep{Uecker}, ADMM-Net~\citep{Yang}, DAGAN~\citep{Yang2}, as well as PISD and nPIDD.
Among them, TV, ADMM-Net, DAGAN, nPIDD were implemented based on the single-channel MRI data, whereas L1-EPSRiT, PIDD, PISD were implemented based on the multi-channel MRI data. The Calgary Campinas multi-channel MRI data was used in this experiment.

The testing results are shown in Table~\ref{tab:comparison}. Zero-filled image (ZF) is undersampled by a Gaussian $30\%$ downsampling trajectory.
According to the comparison study results, the proposed method shows significant improvement for all test indicators compared with other methods. The FID of PIDD is significantly lower compared to PISD and nPIDD. 

Reconstruction samples of six methods, including PIDD, PISD, DAGAN, ADMM-Net, together with GT and ZF, were chosen for unbiased rating by domain experts. PIDD achieved the highest MOS (except GT) both in holistic image and edge information reconstruction.

\begin{table}[htbp]
  \centering
  \caption{Testing results of different method. Results are obtained by Gaussian 2D masks with 30\%. (Bold values indicate the best performed method).}
    \resizebox{\textwidth}{15mm}{
    \begin{tabular}{ccccccccc}
    \toprule
          & \multicolumn{4}{c}{Gaussian 30\%} & \multicolumn{2}{c}{MOS} & \multicolumn{2}{c}{Time} \\
    Method & NMSE  & PSNR  & SSIM  & FID   & Holistic & Edge  & GPU Time (ms)& CPU Time (s)\\
    \midrule
    GT    & —     & —     & —     & —     & 3.83$\pm$0.52 & 3.70$\pm$0.46 & —     & — \\
    ZF    & 0.0247$\pm$0.0013 & 28.4701$\pm$0.3806 & 0.8756$\pm$0.0060 & 103.33 & 1.03$\pm$0.18 & 1.03$\pm$0.18 & —     & — \\
    \midrule
    TV    & 0.0254$\pm$0.0023 & 28.5259$\pm$0.6997 & 0.8840$\pm$0.0117 & 106.04 & —     & —     & —     & 0.64$\pm$0.02 \\
    L1-ESPIRiT & \textbf{0.0071$\pm$0.0008} & 32.4709$\pm$0.8978 & 0.9056$\pm$0.0122 & 21.21 & —     & —     & —     & 53.24$\pm$0.82 \\
    ADMM-Net & 0.0129$\pm$0.0022 & 31.5171$\pm$0.9744 & 0.9368$\pm$0.0115 & 67.97 & 1.10$\pm$0.30 & 1.13$\pm$0.34 & —     & 1.19$\pm$0.12 \\
    DAGAN & 0.0169$\pm$0.0011 & 30.1123$\pm$0.3710 & 0.9072$\pm$0.0053 & 53.67 & 2.60$\pm$0.66 & 2.43$\pm$0.56 & 2.99$\pm$0.01 & 0.21$\pm$0.01 \\
    \midrule
    nPIDD & 0.0099$\pm$0.0009 & \textbf{32.4809$\pm$0.4680} & \textbf{0.9438$\pm$0.0044} & 29.42 & —     & —     & 4.95$\pm$0.01 & 0.47$\pm$0.00 \\
    PISD  & 0.0103$\pm$0.0009 & 32.2685$\pm$0.4808 & 0.9398$\pm$0.0050 & 24.12 & 3.70$\pm$0.59 & 3.26$\pm$0.51 & 4.95$\pm$0.01 & 0.47$\pm$0.00 \\
    PIDD  & 0.0101$\pm$0.0009 & 32.3694$\pm$0.4443 & 0.9425$\pm$0.0046 & \textbf{15.38} & \textbf{3.80$\pm$0.54} & \textbf{3.40$\pm$0.66} & \textbf{4.95$\pm$0.01} & \textbf{0.47$\pm$0.00} \\
    \bottomrule
    \end{tabular}%
    }
  \label{tab:comparison}%
\end{table}%

\begin{figure}[htbp]
\centering
\includegraphics[width=12cm]{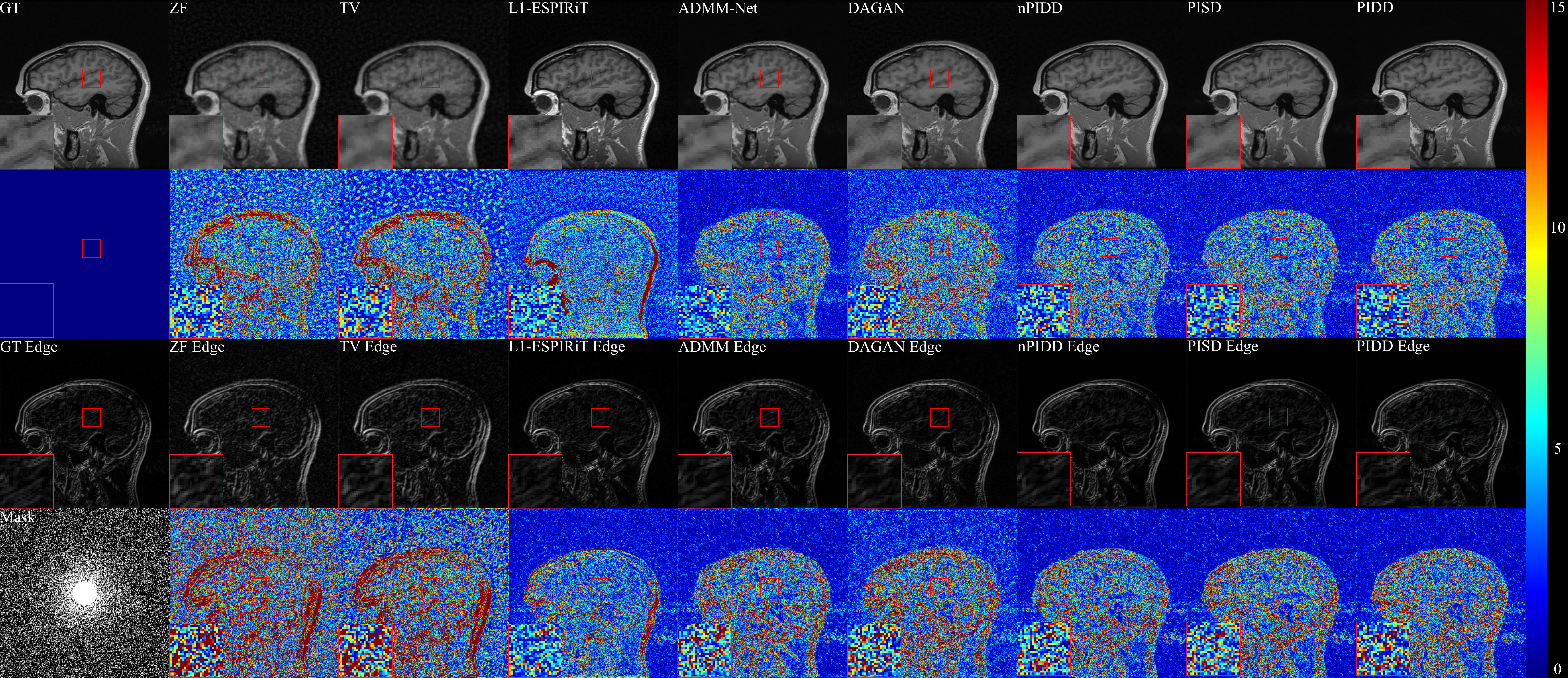}
\caption{Testing examples of different methods. Results are obtained by Gaussian 2D masks with 30\%. Line1: MR images; Line2: Differences of MR images ($\times15$); Line3: Edge Information of MR images; Line4: Differences of Edge Information ($\times15$).}
\label{fig:ComparisonFig}
\end{figure}

Testing examples of different methods are shown in Fig.~\ref{fig:ComparisonFig}. Our method is superior to other methods in terms of overall reconstruction quality and edge information reconstruction. We can see that ZF is quite blurred, mixed with many artefacts. Compared with traditional TV and L1-ESPIRiT methods, the noise reduction of ADMM and DAGAN are greatly improved, but the detailed information is still lost significantly. The reconstruction result of L1-ESPIRiT shows rich details, but it is poor in de-noising and time-consuming. 
Compared with PIDD and PISD, nPIDD lacks multi-channel information. Although the basic structure can be completely restored, and most noise is reduced, the excessive smoothing phenomenon is severe compared to PIDD and PISD. From the zoom-in area, it can be seen that PIDD clearly reconstructs the edge information of the brain, but this structure is very shallow in the results of PISD. 

\begin{figure}[htbp]
\centering
\includegraphics[width=12cm]{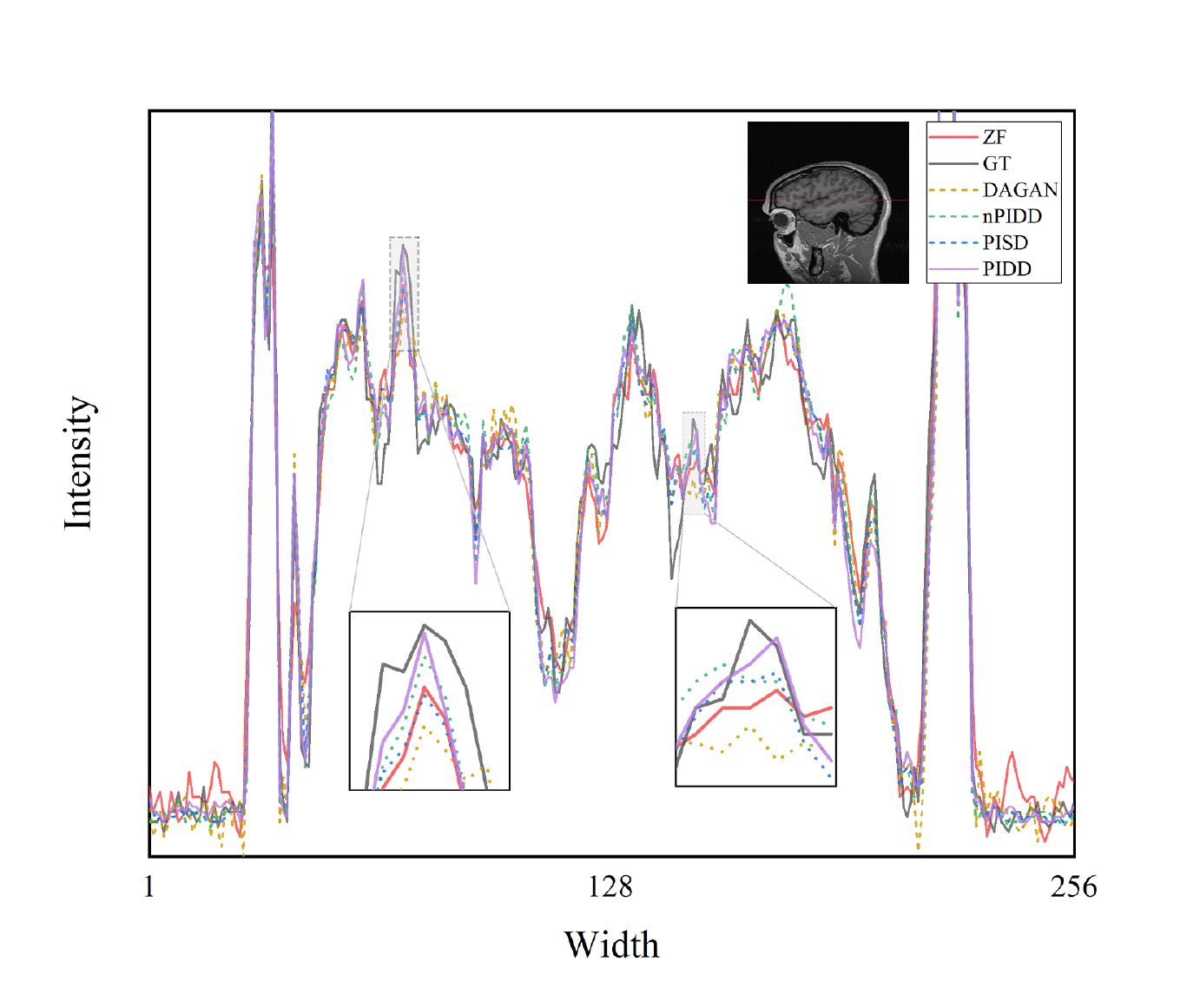}
\caption{Horizontal line profiles of different methods. Results are obtained by Gaussian 2D masks with 30\%.}
\label{fig:ComparisonLineProfile}
\end{figure}

Horizontal line profiles of the samples in Fig.~\ref{fig:ComparisonFig} are shown in Fig.~\ref{fig:ComparisonLineProfile}. ZF and DAGAN still contain lots of noise, while our proposed methods preserve more detail information. Zoom-in areas clearly show that the line profile of PIDD-GAN achieve more accurate than other methods.

\subsection{Experiments on Mask}

In this experiment, different downsampling trajectories were adopted to evaluate the robustness of the proposed method. G2D10\%, G2D20\%, G2D30\%, G2D40\%, G2D50\%, G1D30\% and P2D30\% indicate Gaussian 2D 10\%, Gaussian 2D 20\%, Gaussian 2D 30\%, Gaussian 2D 40\%, Gaussian 2D 50\%, Gaussian 1D 30\% and Poisson 2D 30\% downsampling trajectories, respectively. This experiment was tested on the Calgary Campinas multi-channel MRI dataset.

\begin{figure}[htbp]
\centering
\includegraphics[width=14cm]{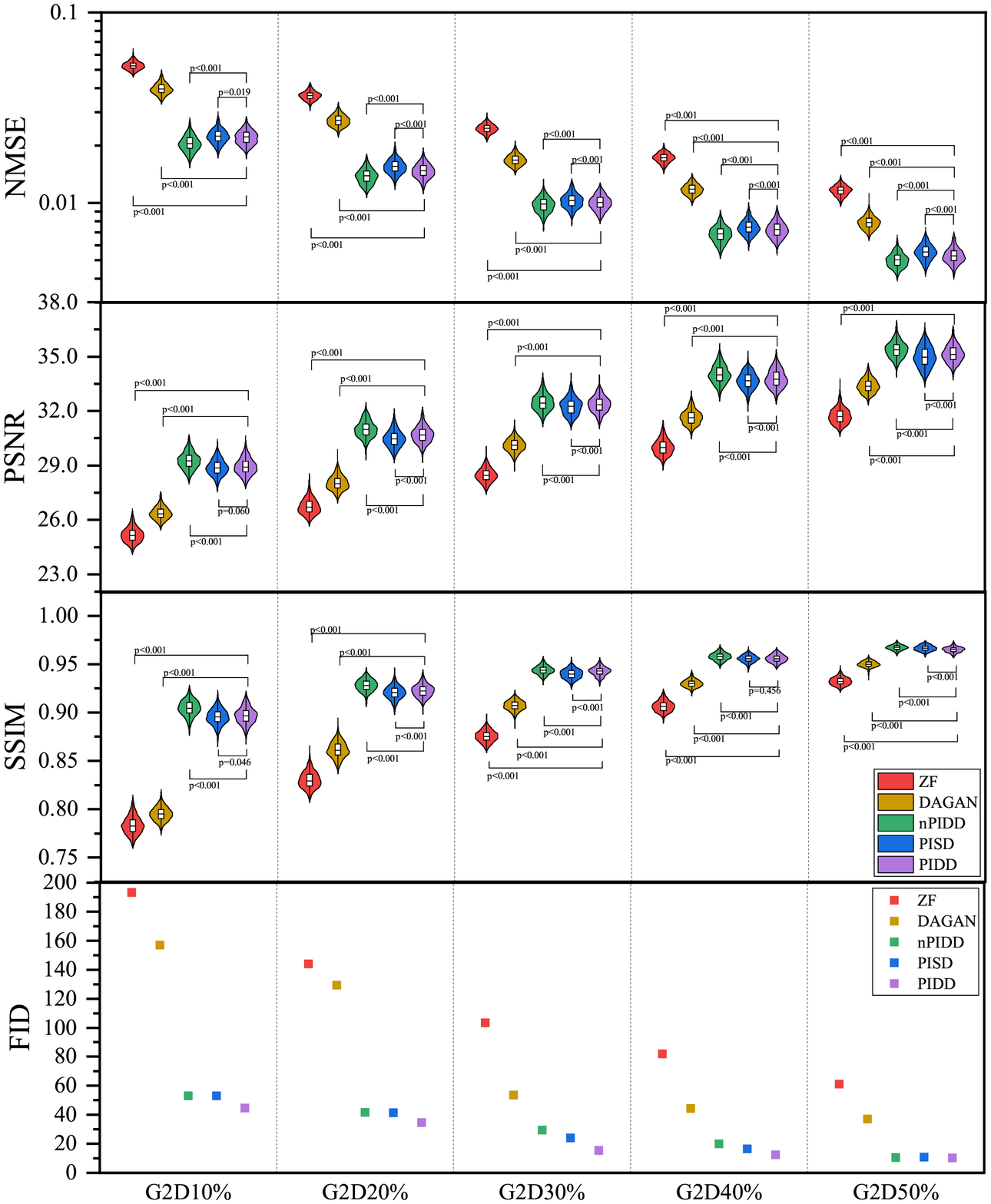}
\caption{Testing results of the experiment on masks using different Gaussian 2D downsampling percentage.}
\label{fig:MASKI}
\end{figure}

\begin{figure}[htbp]
\centering
\includegraphics[width=14cm]{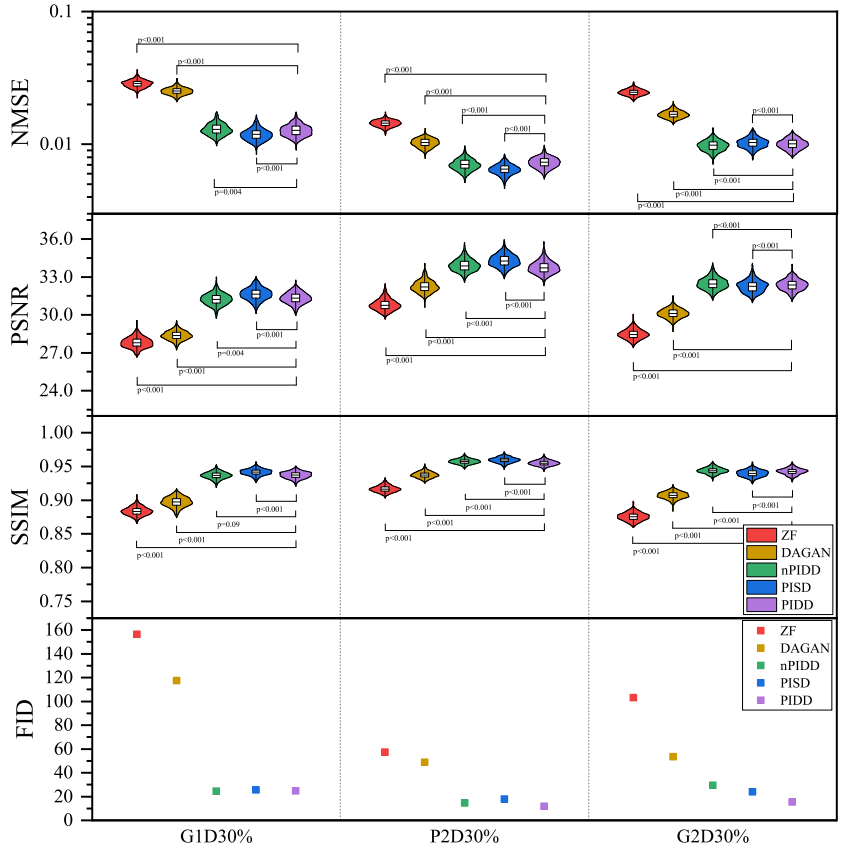}
\caption{Testing results of the experiment on masks using different downsampling trajectories.}
\label{fig:MASKII}
\end{figure}

The testing result are shown in Fig.~\ref{fig:MASKI} and Fig.~\ref{fig:MASKII}.
Experimental results exhibit the same trend under all the different downsampling trajectories. The proposed method provides significant advantages for low downsampling percentage (high acceleration factor), with correspondingly significantly improved reconstruction quality. 

\begin{figure}[htbp]
\centering
\includegraphics[width=12cm]{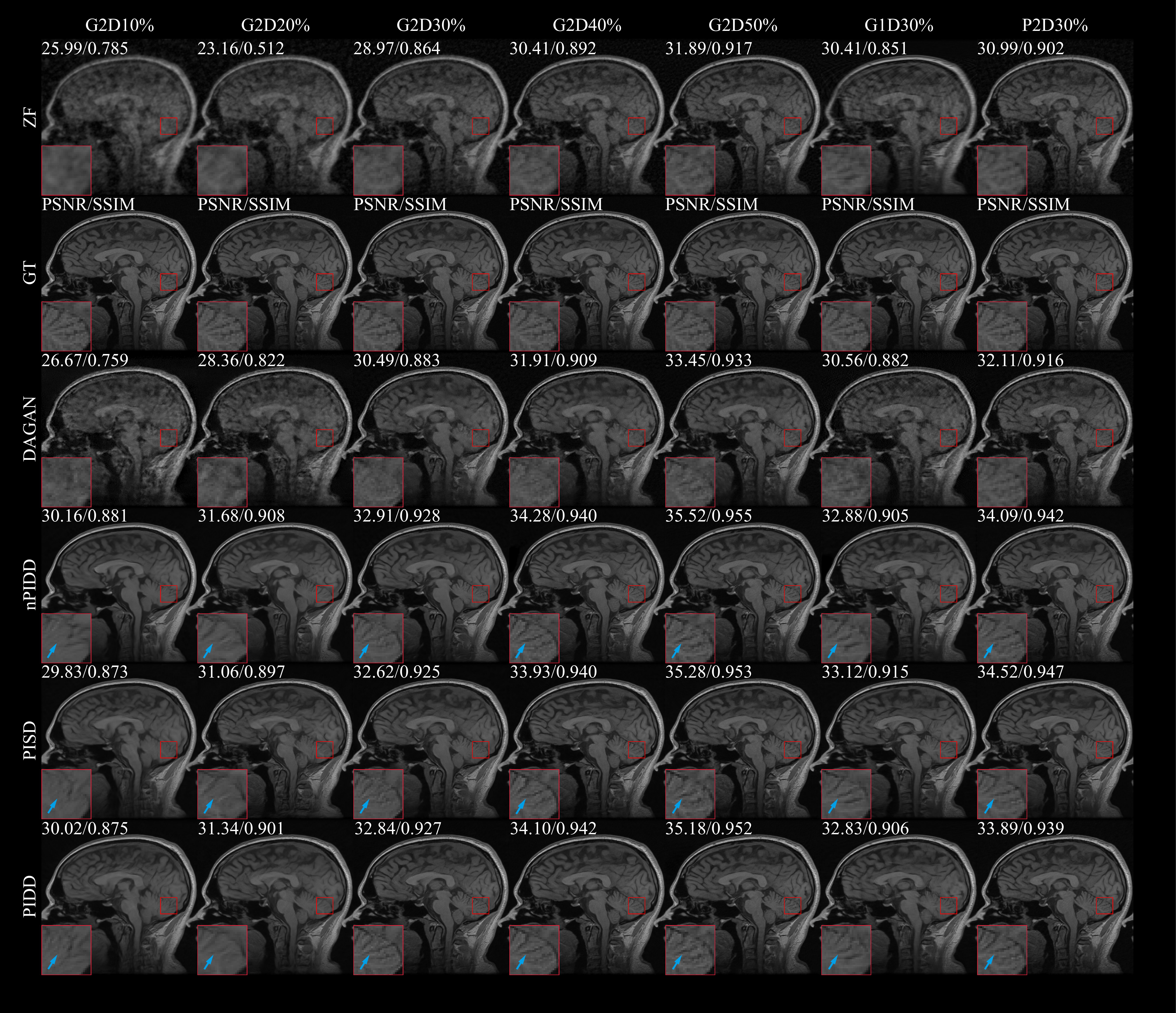}
\caption{Testing examples of the experiment on masks using different downsampling trajectories and percentage.}
\label{fig:ExpMaskFig12}
\end{figure}

The testing examples of the reconstruction are shown in Fig.~\ref{fig:ExpMaskFig12}. 
It can be seen from the results that PIDD has a significant recovery effect on edge information in the case of a low downsampling percentage (10--30\%). The edge information of PIDD, particularly in sulci and cerebellum areas, is greatly preserved, compared to PISD. The texture details of PIDD are also richer than nPIDD. 
In the case of a high downsampling percentage (40--50\%), the reconstruction problem becomes simpler, and the advantages of PIDD with enhanced reconstruction of edge information and the advantages of multi-channel data are less obvious. In the experiment of Gaussian 2D 50\% undersampling, PIDD, PISD and nPIDD basically have the same quality of the reconstruction.

\subsection{Experiments on Noise}

In this experiment, the same downsampling trajectory and different noise levels were used to test the reconstruction performance of the model under the influence of noise. This experiment was tested on the Calgary Campinas multi-channel MRI dataset. Here the noise level ($\text{NL}$) is defined as follows

\begin{align}\label{formula:18}
\text{NL}=\frac{N}{N+S} ,
\end{align}

\noindent where $S$ and $N$ denotes the power of signal and noise respectively. 

\begin{figure}[htbp]
\centering
\includegraphics[width=14cm]{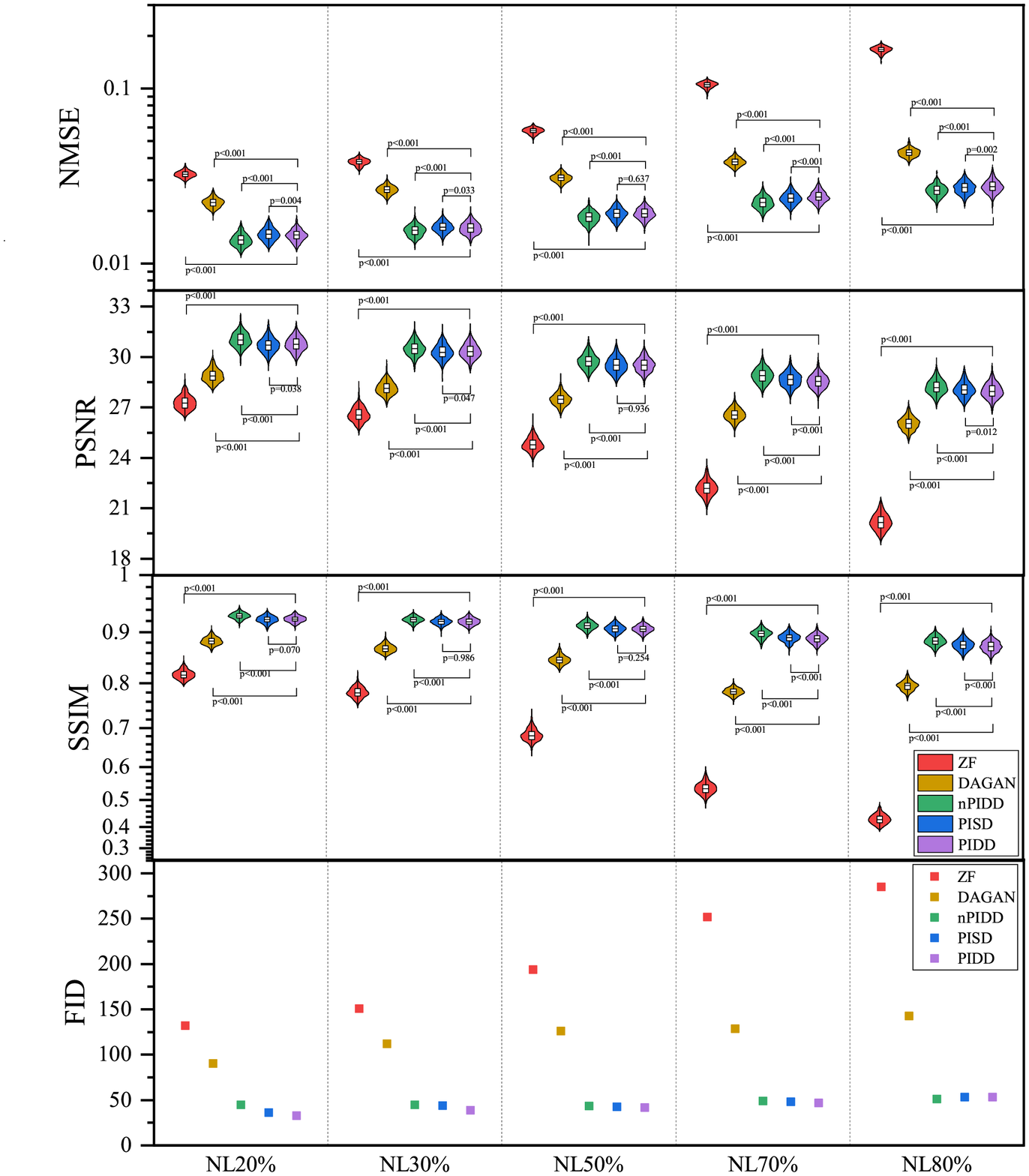}
\caption{Testing results of the experiment on noise using different noise level.}
\label{fig:NOISE}
\end{figure}

\begin{figure}[htbp]
\centering
\includegraphics[width=12cm]{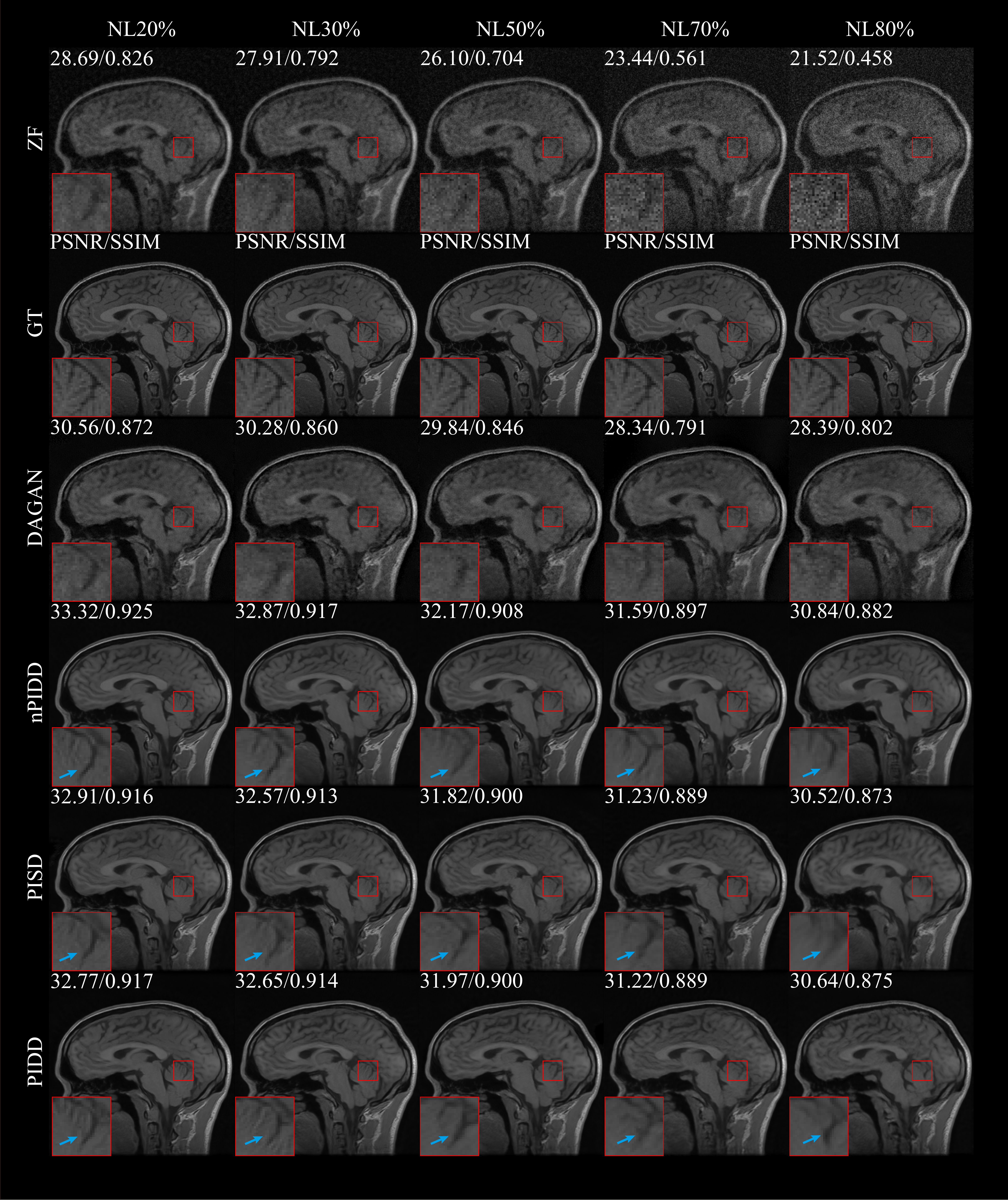}
\caption{Testing examples of the experiment on noise using different noise level.}
\label{fig:ExpNoiseFig10}
\end{figure}

The testing results are shown in Fig.~\ref{fig:NOISE}, and testing samples are shown in the Fig.~\ref{fig:ExpNoiseFig10}.

All considered methods can restore image structure and edge information for low and medium noise levels (20--50\%), with PIDD having strong advantages over PISD and nPIDD. This advantage weakens as noise level increases, and PISD, which focuses on overall information recovery, performance slightly surpasses PIDD, which focuses more on edge information preservation, for high noise (70--80\%).

\subsection{Ablation Experiments on Residual Learning}

In this experiment, the effect of residual learning in the network was discussed. Our proposed model was tested on MICCAI 2013 grand challenge dataset, using Gaussian 1D 30\% downsampling. 

The experiment was divided into four groups: (1) GRLR (model with GR and LR), (2) GRnLR (model with GR without LR), (3) nGRLR (model with LR without GR), (4) nGRnLR (model without LR and GR). Early stopping strategy was turned off in this experiment to prolong the training process for a better and more distinguishable comparison for the training step.

\begin{figure}[htbp]
\centering
\includegraphics[width=20cm, angle=90]{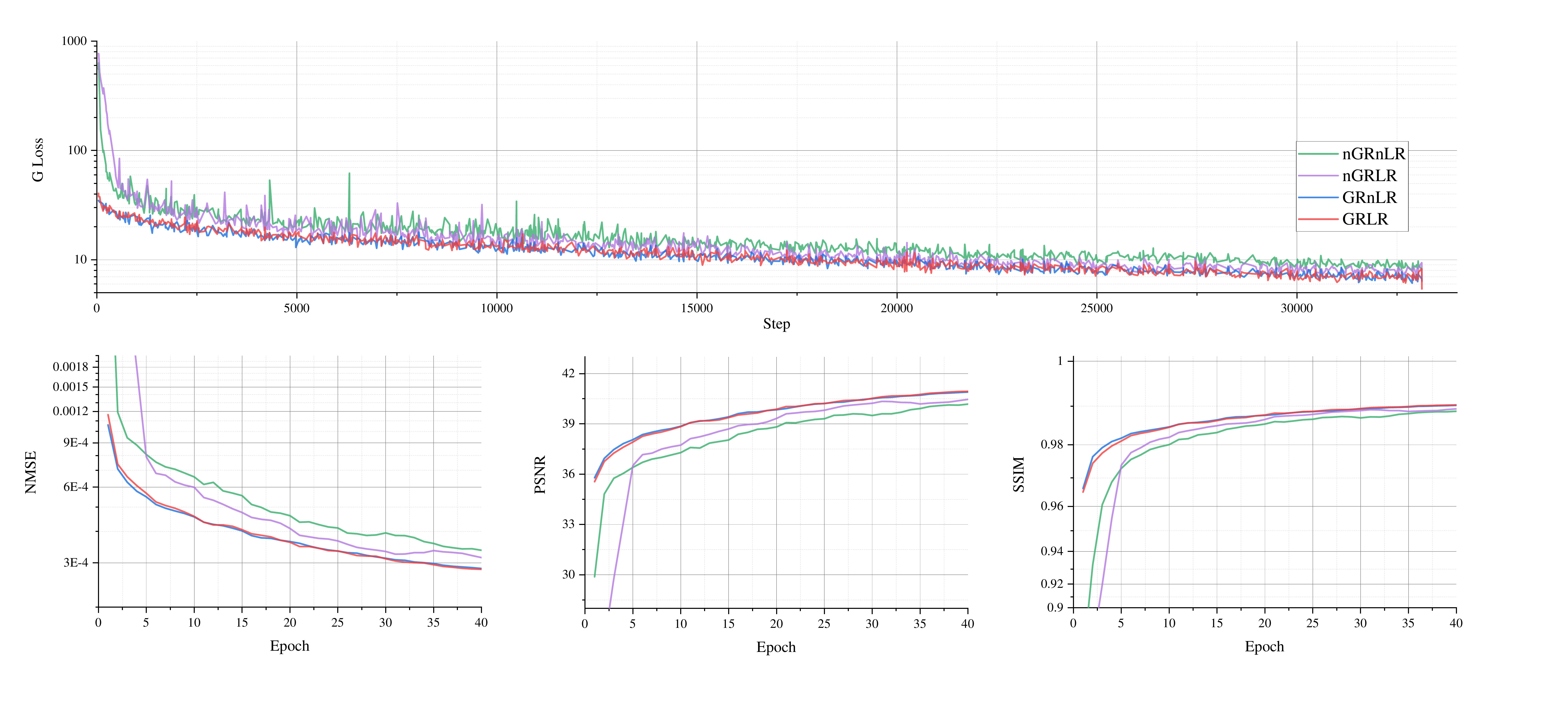}
\caption{Testing results of the ablation studies with residual learning. GRLR: model with GR and LR; GRnLR: model with GR without LR; nGRLR: model with LR without GR; nGRnLR: model without LR and GR.}
\label{fig:ResLearning}
\end{figure}

\begin{figure}[htbp]
\centering
\includegraphics[width=20cm, angle=90]{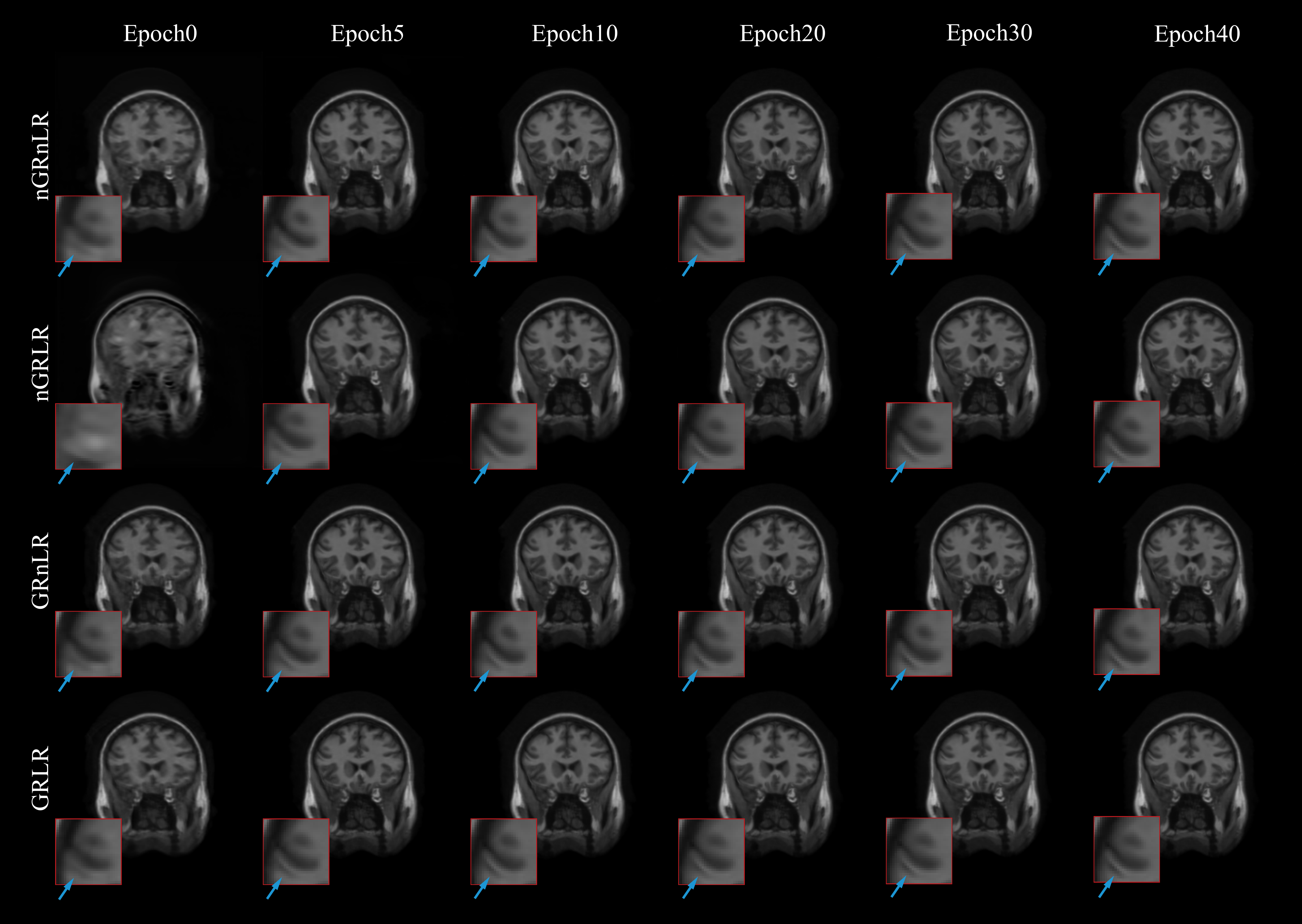}
\caption{Testing examples of the ablation studies with residual learning. GRLR: model with GR and LR; GRnLR: model with GR without LR; nGRLR: model with LR without GR; nGRnLR: model without LR and GR.}
\label{fig:AblExpResFig}
\end{figure}

Fig.~\ref{fig:ResLearning} shows NMSE, SSIM, PSNR and generator loss (G Loss) of the four groups changing with the training process, and Fig.~\ref{fig:AblExpResFig} shows testing examples with respect to different epoch weights. 

Models with GR (GRLR, GRnLR) have faster convergence and better final results compared with those without GR (nGRLR, nGRnLR). If the model applies GR, then using LR has little further impact effect on the results. For non-GR models, nGRLR converges significantly slower than nGRnLR but final results are superior. Therefore, we chose GRLR as the generator for subsequent study.

\section{Discussion}

In this work, we introduce the PIDD-GAN for multi-channel MRI reconstruction using multi-view parallel imaging information, focusing on the enhancement of edge information and the utilisation of the multi-channel MR data.

Experimental results verified that the proposed dual discriminator design does greatly improve reconstruction quality of edge information, particularly for sulci and cerebellum areas with rich edge information. The utilisation of multi-channel data also reduces the excessive smoothing phenomenon to a certain extent, and the texture of the tissue can be also better preserved.

During experiments, we found that SSIM and PSNR do not reflect the reconstruction quality very well, since these two indicators have better tolerance for over-smoothing and are relatively insensitive to the edge information details. Therefore, we adopted FID using a deep network and MOS based on subjective scoring by experts to assess our results more comprehensively.

Meanwhile, we did a series of experiments to test the role of the attention mechanism (FCA Block and SE Block) in the entire model. In most cases, the utilisation of attention mechanisms slightly improved the convergence speed and final results of the model. However, incorporating the attention mechanism degraded the final result for a few cases. The specific mechanism requires further study. 

The proposed model still has some remaining limitations. Dual discriminator structure superiority reduces for high downsampling percentage, and single discriminator structure, focusing on overall recovery, may offer better performance. Noise can affect edge information extraction, particularly for higher noise levels. This effect is very marked when using the Sobel operator, and hence may jeopardise $D_2$ performance, reducing image reconstruction quality.

Further studies will continue to focus using multi-channel information to help edge information reconstruction and restoration. We will also consider how to reduce network calculations and improve network efficiency.

\section{Conclusion}

This work proposed a parallel imaging based dual discriminator generative adversarial network for multi-channel MRI reconstruction, enhancing edge information and multi-channel MR data utilisation using multi-view information. Experiment results verified that the proposed method offers significantly better performance preserving edge information for MRI reconstruction. 

\newpage
\newpage
\section*{Acknowledgement}

This work was supported in part by the Zhejiang Shuren University Basic Scientific Research Special Funds, in part by the European Research Council Innovative Medicines Initiative (DRAGON, H2020-JTI-IMI2 101005122), in part by the AI for Health Imaging Award (CHAIMELEON, H2020-SC1-FA-DTS-2019-1 952172), in part by the UK Research and Innovation Future Leaders Fellowship (MR/V023799/1), in part by the British Heart Foundation (Project Number: TG/18/5/34111, PG/16/78/32402), in part by the Foundation of Peking University School and Hospital of Stomatology [KUSSNT-19B11], in part by the Peking University Health Science Center Youth Science and Technology Innovation Cultivation Fund [BMU2021PYB017], in part by the National Natural Science Foundation of China [61976120], in part by the Natural Science Foundation of Jiangsu Province [BK20191445], in part by the Qing Lan Project of Jiangsu Province, in part by National Natural Science Foundation of China [61902338], in part by the Project of Shenzhen International Cooperation Foundation [GJHZ20180926165402083], in part by the Basque Government through the ELKARTEK funding program [KK-2020/00049], and in part by the consolidated research group MATHMODE [IT1294-19].

\newpage
\newpage
\bibliographystyle{unsrt}
\bibliography{references}

%===============================================================

%\begin{acknowledgements}
%\end{acknowledgements}



\end{document}